\documentclass[preprint]{aastex}
\usepackage{array}
\usepackage{multicol}
\usepackage{amssymb}
\usepackage{epsfig}
\usepackage{ifthen}
\usepackage{rotating}

\newboolean{ispreprint}
\setboolean{ispreprint}{true}

\bibpunct[; ]{(}{)}{;}{a}{}{,}

\newsavebox{\thisbox}
\newlength{\thiswid}

\newcommand{\etal}{\textit{et al.}}

\shorttitle{DECELERATION PARAMETER }
\shortauthors{M.V. JOHN}

\begin{document}

\title{COSMOGRAPHIC EVALUATION OF DECELERATION PARAMETER USING SNe Ia
DATA}

\author{MONCY V. JOHN}

\affil{Department of Physics, St. Thomas College, Kozhencherri, Kerala
689641,
India; }
\email{moncy@iucaa.ernet.in}

\and

\affil{Inter-University Centre for Astronomy and Astrophysics, Pune 411
007,
India;}

\begin{abstract}
The apparent magnitude-redshift data of SNe Ia call for modifications in
the standard model energy densities. 
Under the circumstance
that this modification cannot be limited to the addition of a mere
cosmological
constant, a serious situation has emerged in cosmology, in which 
 the energy densities in the universe have become largely
speculative. In this situation, an equation of state of the form
$p=w\rho$ itself
is not well-motivated. In this
paper, we argue that the  reasonable option left is to make a
model-independent
analysis of SNe data, without reference to the energy densities. In this
basically
kinematic approach, we limit ourselves to the observationally justifiable
assumptions of homogeneity and isotropy; i.e., to the  assumption that
the 
universe
has a RW metric. This cosmographic approach is historically the original
one
to cosmology. We perform the analysis by expanding the scale factor into
a
polynomial of order 5, which assumption can be further generalised to any
order.
The present expansion rates $h$, $q_0$, $r_0$ etc. are evaluated by
computing
the marginal likelihoods for these parameters. These values are relevant,
since any
cosmological solution would ultimately need to explain them.

\end{abstract}


\section{INTRODUCTION}
 In the
Friedmann models of the universe, the solution
of the
Einstein's field equation requires an equation of state to relate
the energy density and the
pressure of the universe \citep{wein}.  An important recent development
in
cosmology is that 
the  apparent magnitude-redshift data of SN Ia
\citep{rie98,per99,riess1,perl1} give reasons to suspect that the
expansion
of 
the universe is accelerating and thereby points to a major gap in our
understanding of
the density components in the universe and the equations of state obeyed
by
them.  In fact, this gap leaves the solution of the Einstein
equation speculatory to a great
extent.  The explanation of all other cosmological observations needs
this
solution, as it describes the expansion of the background spacetime.
Hence
these SN data, which are claimed to be the only qualitative signature of
an
accelerated expansion of the universe \citep{riess1}, can be
said to put
cosmologists back to square one.  An important question we need to face
at
the
outset is 
whether we view this suspected accelerated expansion  as a new 
phenomenon exhibited by the universe in the 
recent epoch  or as a caveat in our
understanding of the dynamics of the universe. 
In this paper, we adopt the position that whereas there is general
agreement that the theory
needs modifications for the present epoch, theories and observations
regarding the past  require more careful scrutiny and hence
while
analysing these data,
 dependence on specific models needs to be avoided.

Nevertheless, the traditional approach to cosmology is through
cosmography \citep{wein}, rather than
through
the cosmological solutions.  For the determination of the Hubble
parameter
$H_0$, this method is used even now, and the value is considered reliable
when
one uses only low redshift objects.  In order to measure $q_0$, the
present
value
of the cosmic deceleration parameter, it needs to go to higher redshift
objects
(such as the  brightest galaxy clusters or supernovae). Weinberg
(1972)
discusses various attempts made in the past in this direction at length.
Those
measurements used the brightest cluster galaxies as distance indicators,
which are
poor in reliability, and this led Weinberg to remark that, about the
precise
value
of $q_0$, the knowledge (in 1972) is as little as that in 1931.

It seems that this interest in the value of $q_0$ declined, as
cosmologists
became complacent with the standard model (which is  a Friedmann model
with either dust or radiation as component).  Even after the release of
substantial amount of SN Ia data in recent years, though it is generally
believed that these data imply an  accelerating universe, only very few
works are done to evaluate the cosmic deceleration parameter in any truly
model-independent way, i.e., without making any assumptions on the energy
densities etc.  Instead, most analyses of SN Ia data aim at defending the
standard cosmological solutions or their variants, which are obtained
 by way of introducing exotic energy densities
with strange equations of state and so on. When there is no clear
prescription for $\rho$, the density of the universe, even an equation of
state of the form $p=w\rho$ is not well-motivated. It is unfortunate that
some of
the SN data available in the literature itself is in a form suitable 
only to specific models.

However, some exceptions to this trend exist \citep{q1,q2,q3, q4}. 
Trentham
(2001)
examines the effects of various assumptions we make in evaluating the
cosmic
acceleration, such as the large scale isotropy and homogeneity, flatness,
etc.  and
concludes that failures of two or more such assumptions in concordance
may
have
strong effects.  Daly and Djorgovski (2003) aim at a model-independent
way to
compute the deceleration parameter $q$ as a function of redshift $z$.
They  divide the data points into
redshift intervals for which a dimensionless coordinate distance $y(z)$
is
fitted in the form of second order polynomials.  These are then
numerically
differentiated to obtain $q(z)$.  An attempt is also made to find the
value of
the redshift at which the speculated transition from deceleration to
acceleration occurred.
 Wang and Mukherjee (2004) allow the dark energy density to be an
arbitrary
function of the redshift $z$ and present a model independent
reconstruction
of the dark energy density and have found that it varies with time at less
than $2\sigma$.

This paper aims at evaluating $q_0$ cosmographically, without making any
kind of
assumptions regarding its energy content. The method 
requires the least amount of speculatory inputs in the analysis of the
supernova data.
 We confine ourselves to the
assumptions that
the universe has a Robertson-Walker (RW) metric with $k= \pm 1$
or 0
and that the scale factor can be approximated by the first six terms in
a
Taylor series; i.e., as a fifth order polynomial. (In \citep{wein}, only
the first three terms are taken. As we shall see later, the complexity
increases  with the number of terms taken, which will require
 high computing power to deal with.) The
coefficients in this expansion are considered as
the parameters of the theory.  The only other parameter, which does not
come
into this
expansion is  $M$, the absolute luminosity of the standard candles. We
evaluate
the likelihood for these
parameters for all possible combinations of parameter values. The data 
used are that of the 54 `All SCP SNe' reported in \citep{perl1}.  For the
parameters of interest, we find the marginal likelihood by
marginalising over all other parameters. This method has the advantage
that
while giving the likelihood for one parameter, it takes care of the
uncertainties
in  the other ones.

The method is simple and is a straight-forward generalisation of the one
discussed
in \citep{wein}.
What is presented here are some first results, which can be improved
substantially with the fastness of computation.  This method, which can
be
described as basically kinematic, gives
valuable information regarding the expansion rates of the universe and
provides
a testing ground for all cosmological solutions.

\section{THE METHOD}

To start with, we state our assumption explicitly: the universe has a RW
metric, with
$k=\pm 1$ or 0. Then the scale factor can be
expanded into a Taylor series

\begin{equation} \label{eq:a1}
a(t)=a_0\left[
1+\frac{a^{(1)}_0}{a_0}(t-t_0)+\frac{a^{(2)}_0}{2!a_0}(t-t_0)^2+
\frac{a^{(3)}_0}{3!a_0}(t-t_0)^3+ \frac{a^{(4)}_0}{4!a_0}(t-t_0)^4 +
\frac{a^{(5)}_0}{5!a_0}(t-t_0)^5 + ...
  \right].
\end{equation}
Henceforth, we limit ourselves by keeping only up to  the fifth order
term in the above series and assume that it is a good approximation. Here
$a^{(i)}_0$ refers
to the present value of the $i^{th}$ derivative of the
scale factor $a$ with
respect to time. We substitute

$$
\frac{a^{(1)}_0}{a_0}\equiv H_0 \equiv a_{(1)},
\qquad
\frac{a^{(2)}_0}{2!a_0}\equiv -\frac {q_0H_0^2}{2!} \equiv a_{(2)},
\qquad
\frac{a^{(3)}_0}{3!a_0}\equiv \frac {r_0H_0^3}{3!} \equiv a_{(3)},
\qquad
\frac{a^{(4)}_0}{4!a_0}\equiv -\frac {s_0H_0^4}{4!} \equiv a_{(4)}, 
$$
\begin{equation}
\frac{a^{(5)}_0}{5!a_0}\equiv \frac{u_0H_0^5}{5!} \equiv a_{(5)}, 
\label{eq:pardef}
\end{equation}
and

\begin{equation}
\label{eq:T}t-t_0\equiv T
\end{equation}
to get

\begin{eqnarray}
a(t_0+T)&=&a_0\left[
1+H_0T-\frac{q_0H_0^{2}}{2!}T^2+
\frac{r_0H_0^3}{3!}T^3-  \frac{s_0H_0^4}{4!}T^4 +\frac{u_0H_0^5}{5!}T^5
\right]  \nonumber \\
&=& a_0\left[1+a_{(1)}T+a_{(2)}T^2+a_{(3)}T^3+a_{(4)}T^4+a_{(5)}T^5\right]
\label{eq:a2}.
\end{eqnarray}
Now one can  write

\begin{equation}
\label{eq:a3}\frac{1}{a(t_0+T)}=\frac{1}{a_0}\left(
1+\beta T+\gamma T^2+
\delta T^3+\epsilon T^4 + \kappa T^5+ \mu T^6 + ... \right),
\end{equation}
where

$$
\beta =-a_{(1)},
\qquad
\gamma = -a_{(1)}\beta -a_{(2)},
\qquad
\delta = -a_{(1)} \gamma -a_{(2)}\beta -a_{(3)},
\qquad
\epsilon = -a_{(1)}\delta -a_{(2)}\gamma -a_{(3)}\beta -a_{(4)},
$$
$$
\kappa =-a_{(1)}\epsilon -a_{(2)}\delta -a_{(3)}\gamma -a_{(4)}\beta
-a_{(5)}
$$
and

\begin{equation}
\mu =-a_{(1)}\kappa -a_{(2)}\epsilon -a_{(3)}\delta -a_{(4)}\gamma
-a_{(5)}\beta, \qquad
\hbox{etc.}
\end{equation}
Even though our assumption is that  of a fifth order polynomial for the
scale factor, the series (\ref{eq:a3}) cannot be terminated anywhere
without
proper
checking of the remainder term. We have kept terms up to order 6 in T for
better
accuracy. This can still be improved if the need arises.

For a $k=0$ RW metric,

\begin{equation}
r_1= \int_{t_1}^{t_0} \frac{cdt}{a(t)}= \frac{c}{a_0}\int_{T_1}^0 \left(
1+\beta T+\gamma T^2+
\delta T^3+\epsilon T^4 + \kappa T^5+ \mu T^6 + ... \right) dT
\end{equation}
(where $c$ is the velocity of light), or

\begin{equation}
\label{r1a0k0}r_1a_0 = c\left( -T_1-\frac{\beta T_1^2}{2} - \frac{\gamma
T_1^3}{3}-\frac{\delta T_1^4}{4} - \frac{\epsilon T_1^5}{5} -
\frac{\kappa
T_1^6}{6}-  \frac{\mu T_1^7}{7} - .. \right).
\end{equation}
Similarly for the  $k=\pm 1$ cases, we have

\begin{equation} \label{r1a0k+1}r_1a_0 = a_0\; S_k \left[
\frac{c}{a_0}\left(
-T_1-\frac{\beta T_1^2}{2} -\frac{\gamma T_1^3}{3}-\frac{\delta T_1^4}{4}
-
\frac{\epsilon T_1^5}{5} - \frac{\kappa T_1^6}{6} - \frac{\mu T_1^7}{7} -
.. 
\right) \right], 
\end{equation}
where $S_k(x)=\sin x$ for $k=+1$ and $S_k(x)=\sinh x$ for $k=-1$.
However,
since the $k=0$ case is identical to $a_0 \gg 1$, we consider only the
$k=\pm 1$ cases. We invert the following equation

\begin{equation}
 1+z=\frac{a(t_0)}{a(t_0+T_1)}=1+\beta T_1+\gamma T_1^2+ \delta
T_1^3+\epsilon T_1^4 + \kappa T_1^5 + \mu T_1^6 + ...  \label{eq:numsolzT}
\end{equation} 
numerically to obtain $T_1$ in terms
of $z$ and substitute in the above equation for $r_1a_0$ and compute

\begin{equation}
D=r_1a_0(1+z),
\end{equation}

\begin{equation}
m=5 \log \left(\frac{D}{1 \hbox {Mpc}}\right)+25+M.
\end{equation}
Here $D$ and hence $m$ are functions of $z$, $a_0$, $\beta$ , $\gamma$ ,
$\delta$ , $\epsilon $, $\kappa$, $\mu $ and $M$ (or equivalently $z$,
$a_0$,
$H_0$,
$q_0$,
$r_0$, $s_0$, $u_0$, $k$, and $M$).

The likelihood function is

\begin{equation}
{\cal L}= \exp(-\chi^2(a_0,h,q_0,r_0,s_0,u_0,k,M)/2).
\end{equation}
The marginal likelihood $L$ \citep{drell,rie98,mvjvn} for each of $h$,
$q_0$, $r_0$, $s_0$ and $u_0$ are found by assuming flat
priors for the remaining parameters in the set $a_0$, $h$, $q_0$, $r_0$,
$s_0$, $u_0$, $k$, and $M$, in the corresponding intervals
$\Delta a_0$, $\Delta h$, $\Delta q_0$, $\Delta r_0$, $\Delta s_0$,
$\Delta u_0$, $\Delta
k$, and $\Delta M$
as the case may be. For example, $L(q_0)$ is found as

\begin{equation}
 L(q_0) = \frac{1}{2}\sum_{k=-1,1}\frac{1}{\Delta
a_0}\frac{1}{\Delta h}\frac{1}{\Delta r_0}\frac{1}{\Delta s_0}
\frac{1}{\Delta u_0} \frac{1}{\Delta
M} \int
e^{-\chi^2 /2} da_0\; dh\; dr_0\; ds_0\; du_0\; dM.
\end{equation}
The factor outside the integral is the product of flat priors. For our
purpose,
the value of this  is uninteresting.

\section{COMPUTATION OF LIKELIHOOD FUNCTIONS}

The data  analysed are that found in \citep{perl1}, in Tables 3, 4, and 5
and
are reproduced in Table 1 of this paper. These 54 SNe
Ia belong to the subset `All SCP SNe Ia', mentioned in Table 8 of
\citep{perl1}. But the data we are using are slightly different from 
that
used
by these authors for
cosmological fits. The apparent magnitudes  used here are stretch
corrected,
which relieves us of the one parameter $\alpha$ in the calculations.

An important part of the computation is the numerical integration to
obtain the
marginal likelihood. Another part is the numerical solution of equation
(\ref{eq:numsolzT}) for the cosmic time $T_1$, given the redshift $z$,
for
each
set of parameter values.  The results for these numerical solutions are
cross-checked and found
that they are reliable. However, provisions were left to discard those
parameter
values for which the accuracy is poor (i.e., when the solution for $T_1$
is
such that the difference between the left and right sides of the equation
is greater than 0.001).
In another part, we find the reciprocal of the polynomial for $a(t)$ [see
equation (\ref{eq:a3})]  numerically. This procedure is prone to more
serious
errors due to the truncation of the series. In order to minimise it, we
have kept seven terms on the right hand
side.
We cross-check also this calculation and discard those parameter values
which
do not give sufficient accuracy for $1/a(t)$, as in the previous case. 

The resulting curves $L$ are shown in   Figures 1 - 4. The
mean
value of $q_0$ obtained from Figure 2 is  $<q_0> \approx -0.77$.
  Here
the
ranges
of
the parameters used in the integral are $3000<a_0<8000$, $0.6<h<0.8$,
$-15<r_0<15$, $-65<s_0<65$, $-150<u_0<150$, $-1<k<1$, and $-19.6<M<-19.1$,
where
the step sizes are
$\delta a_0=1000$, $\delta h=0.04$, $\delta r_0=0.5$, $\delta s_0=1$,
$\delta
u_0=25$,
$\delta k=2$, and $\delta M=0.1$. The parameter $M$ has reasonable 
prior,
which
we have employed in the above case. The prior range $a_0 > 3000$ Mpc is
also
reasonable and
the
upper limit (8000 Mpc) is chosen high enough to incorporate flat ($k=0$)
models
indirectly. In order to ascertain the validity of the priors for the
parameters
$h$, $r_0$, and $s_0$, the marginal likelihoods for these parameters are
computed and are given in Figures 1, 3 and 4. These curves show that the
contributing
regions of these parameters are well within the ranges we have employed.

The mean values of the other parameters are the following: $<h>\approx
0.67$, 
$<s_0>\approx -2.67$, and $<r_0>\approx
2.64$. The
marginal likelihood for $h$ clearly shows that it is not possible to
narrow
down the estimation of this parameter, even with such refined and high
redshift
data.

\ifthenelse{\boolean{ispreprint}}{
\begin{figure*}[p]
\begin{lrbox}{\thisbox}
\epsfig{file=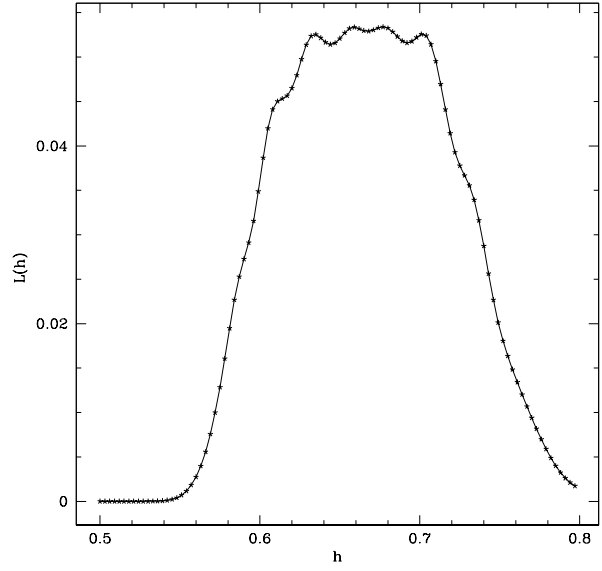, width=3.26in}
\end{lrbox}
\settowidth{\thiswid}{\usebox{\thisbox}}
\begin{center}
\usebox{\thisbox}
\caption{The marginal likelihood for  $h$.}
\label{fig:hlik}
\end{center}
\end{figure*}
}{\placefigure{fig:hlik}}

\ifthenelse{\boolean{ispreprint}}{
\begin{figure*}[p]
\begin{lrbox}{\thisbox}
\epsfig{file=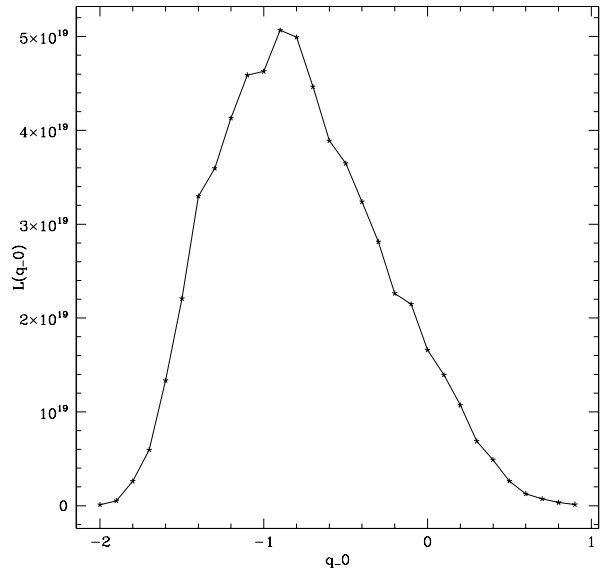, width=3.26in}
\end{lrbox}
\settowidth{\thiswid}{\usebox{\thisbox}}
\begin{center}
\usebox{\thisbox}
\caption{The marginal likelihood for $q_0$.}
\label{fig:qlik}
\end{center}
\end{figure*}
}{\placefigure{fig:qlik}}

\ifthenelse{\boolean{ispreprint}}{
\begin{figure*}[p]
\begin{lrbox}{\thisbox}
\epsfig{file=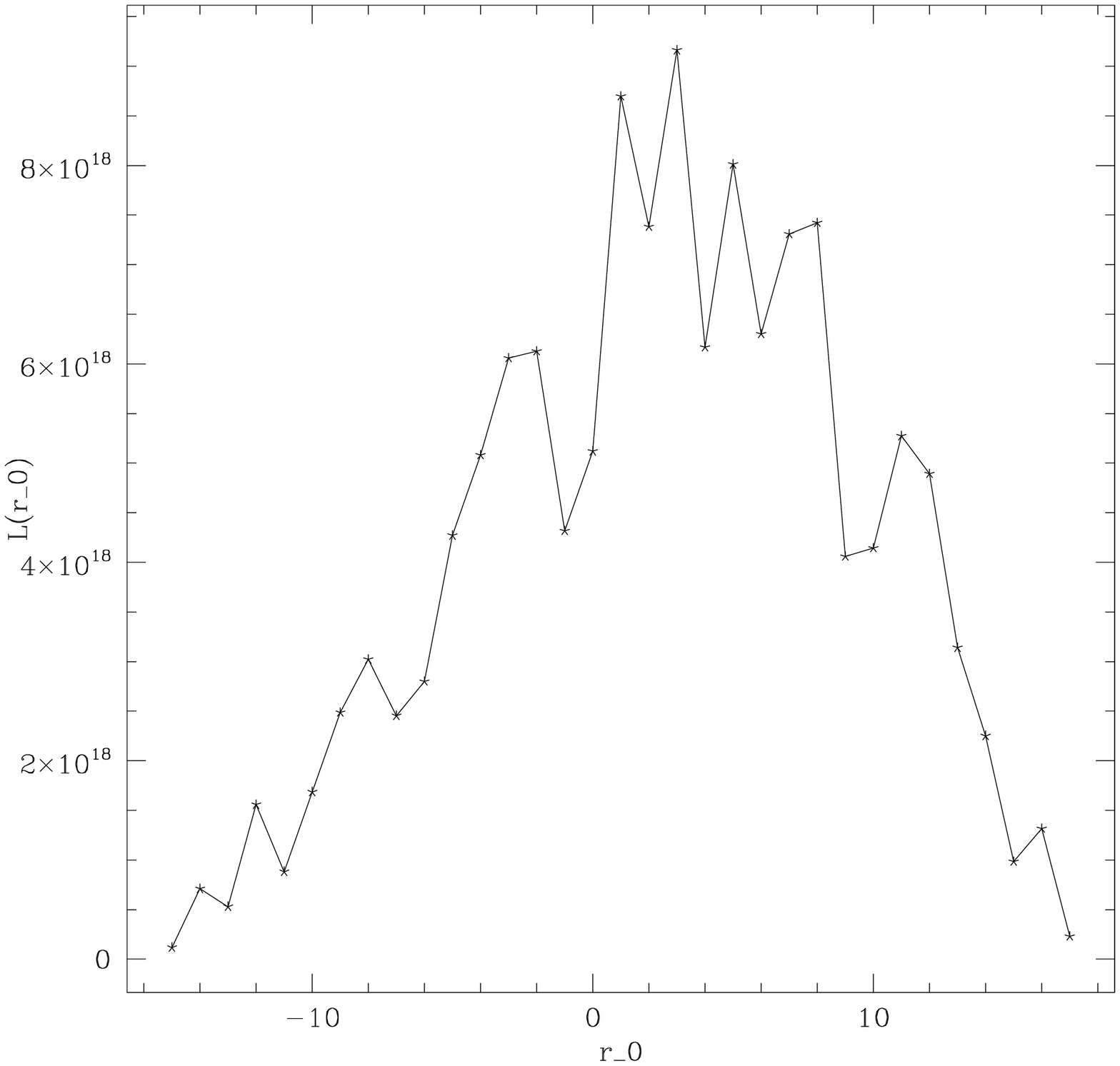, width=3.26in}
\end{lrbox}
\settowidth{\thiswid}{\usebox{\thisbox}}
\begin{center}
\usebox{\thisbox}
\caption{The marginal likelihood for the parameter $r_0$.}
\label{fig:rlik}
\end{center}
\end{figure*}
}{\placefigure{fig:rlik}}

\ifthenelse{\boolean{ispreprint}}{
\begin{figure*}[p]
\begin{lrbox}{\thisbox}
\epsfig{file=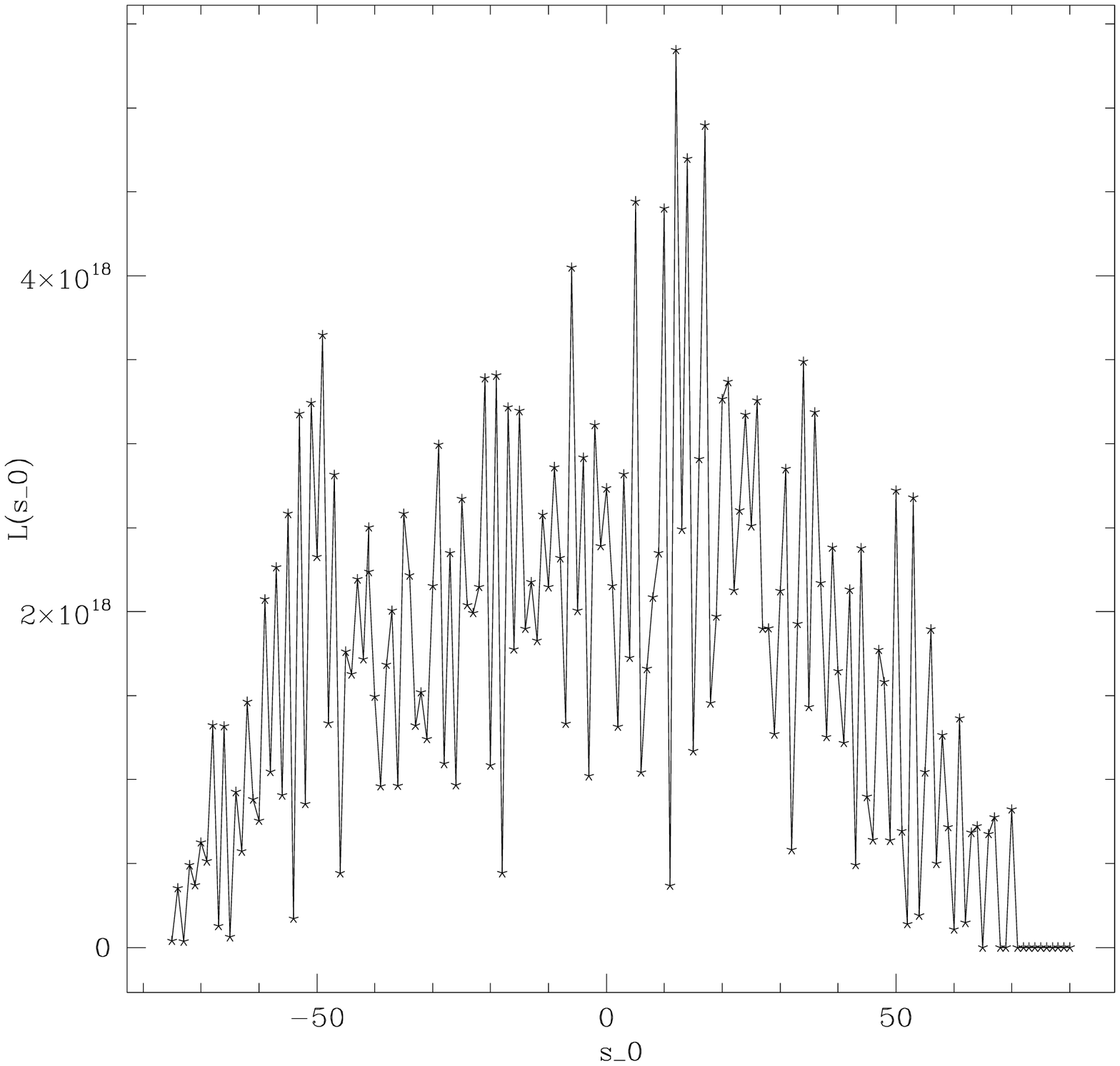, width=3.26in}
\end{lrbox}
\settowidth{\thiswid}{\usebox{\thisbox}}
\begin{center}
\usebox{\thisbox}
\caption{The marginal likelihood for the parameter $s_0$.}
\label{fig:slik}
\end{center}
\end{figure*}
}{\placefigure{fig:slik}}

\section{DISCUSSION}

What we present here is just the revival of an old method of
analysing
the apparent magnitude-redshift data. With the availability of modern
high
performace computers, this becomes a natural choice. Under the
circumstance
that there is no clear prescription for the kind and number of energy
densities
present and no definite motivation for an equation of state of the form
$p=w\rho$, perhaps, 
the 
only option left to understand the dynamics of the universe from SNe data
is  to
evaluate the expansion rates such as  $H_0$, $q_0$, $r_0$, etc., as we
have
done here. The rationale behind the method is straight-forward and
well-established. It is limited only for the perfection of the
computations 
involved. The  calculation of $L(q_0)$ took nearly 75 hours on
using a Pentium 4 Linux machine. However, the accuracy can still be
improved
with the
use of more advanced
computational
facilities. 

A potential source of error in the present cosmographic approach is that
the
perturbative expansion of a(t) used in this paper necessitates
truncations; these result in the breaking down of the approximations
for certain parameters, which are then left out. This leads to
fluctuations in the likelihood functions of higher  derivative expansion
rates
like $r_0$ and $s_0$ (see Figs. 3-4). It is found that if we  use  less
number of terms in the perturbative expansion (for example, a fourth order
polynomial for $a(t)$), even $L(q_0)$  exhibits some gaps, eventhough its
envelope
and the mean value  $<q_0>\approx -0.76$ calculated therefrom is nearly
the
same as that in the present case ($<q_0>\approx -0.77$).  Thus it is
reasonable to expect that the fluctuations in $L(r_0)$ and $L(s_0)$ in
Figures
3 and 4 will disappear, while their envelopes and mean values may remain
mostly
unchanged, if   higher order polynomials than that used in this paper are
employed. (But it should be noted that the addition of one more term will
increase
the computation time at least by an order of magnitude.) However, since
our
primary concern is the evaluation the deceleration parameter $q_0$, we
only
want
to ascertain that the contributiong regions of $r_0$, $s_0$ etc.  are
$-15<r_0<15$, $-65<s_0<65$ etc., which regions we have chosen in our
calculations. A similar exercise is performed also for $u_0$. It is thus
presumed that  Figures 3 - 4 serve their purpose.

Any error in the numerical integration can be reduced by
decreasing the step sizes $\delta q_0$, $\delta h$, etc. While finding
$1/a(t)$
in equation (5), we can add more terms to the right hand side to get more
accurate results. In the numerical solution of equation (10) for $T_1$ in
terms
of $z$ by the Newton-Raphson method, the number of iterations may be
increased,
but even in the present
case, it does not need to go beyond $i=10$.

This method is complementary to the other  analyses of SN
data mentioned in the introduction. The  
complementarity of this approach to those of others arises from the fact
that it is model-independent and perturbative.
The advantage of  evaluating the marginal
likelihood function is that it automatically takes care of the
uncertainties
in 
the other parameters.
  (The
marginalisation technique  used is the same as that used to eliminate
the
combination of Hubble parameter and absolute magnitude of SN Ia by
Riess et.al. \citep{rie98}). Also, while evaluating $L(q_0)$ and other
likelihoods, we use the entire
SN data set to estimate
 the value of the single parameter at the present epoch, and hence our
evaluation of $q_0$ can be considered  reliable.

Our cosmographic method encompasses all  models which have the RW metric
and
hence it is more general than the ones used for Friedmann models. The
evaluation of $L(q_0)$ is
not by restricting ourselves to any specific equation of state or to the
number
of choices of different possible energy densities. The mean value of
$q_0$
 shows that the universe is  indeed accelerating. Evaluation of all the
expansion rates like $q_0$, $r_0$, etc. is important because it would 
ultimately become mandatory for every
viable cosmological models to explain the  values we have obtained for
these
parameters.

\acknowledgments
It is a pleasure to thank Professor J. V. Narlikar and Professor K. Babu
Joseph for several useful discussions.

\begin{table}[H]
\scriptsize\renewcommand{\arraystretch}{1.0}
\begin{lrbox}{\thisbox}
\begin{tabular}{lccc}
\tableline
\tableline
SN & z & m & error \\
\tableline
1997ek &0.863 & 24.59 &  0.19 \\
1997eq &0.538 & 23.15 &  0.18\\
1997ez &0.778 & 24.41 &  0.18\\
1998ay &0.638 & 23.92 &  0.19\\
1998ba &0.430 & 22.90 &  0.18\\
1998be &0.644 & 23.64 &  0.18\\
1998bi &0.740 & 23.85 &  0.17\\
2000fr &0.543 & 23.16 &  0.17\\
1995ar &0.465 & 23.35 &  0.22\\
1995as &0.498 & 23.74 &  0.23\\
1995aw &0.400 & 22.57 &  0.18\\
1995ax &0.615 & 23.38 &  0.22\\
1995ay &0.480 & 22.90 &  0.19\\
1995az &0.450 & 22.66 &  0.20\\
1995ba &0.388 & 22.60 &  0.18\\
1996cf &0.570 & 23.30 &  0.18\\
1996cg &0.490 & 23.11 &  0.18\\
1996ci &0.495 & 22.78 &  0.18\\
1996cl &0.828 & 24.49 &  0.46\\
1996cm &0.450 & 23.11 &  0.18\\
1997F  &0.580 & 23.57 &  0.20\\
1997H  &0.526 & 23.09 &  0.19\\
1997I  &0.172 & 20.29 &  0.17\\
1997N  &0.180 & 20.48 &  0.17\\
1997P  &0.472 & 22.99 &  0.18\\
1997Q  &0.430 & 22.52 &  0.17\\
1997R  &0.657 & 23.80 &  0.19\\
1997ac &0.320 & 21.96 &  0.17\\
1997af &0.579 & 23.38 &  0.18\\
1997ai &0.450 & 22.63 &  0.22\\
1997aj &0.581 & 23.16 &  0.18\\
1997am &0.416 & 22.63 &  0.18\\
1997ap &0.830 & 24.38 &  0.18\\
1990O  &0.030 & 16.33 &  0.20\\
1990af &0.050 & 17.39 &  0.18\\
1992P  &0.026 & 16.14 &  0.19\\
1992ae &0.075 & 18.35 &  0.18\\
1992al &0.014 & 14.42 &  0.23\\
1992aq &0.101 & 19.12 &  0.17\\
1992bc &0.020 & 15.18 &  0.20\\
1992bg &0.036 & 16.66 &  0.20\\
1992bh &0.045 & 17.64 &  0.18\\
1992bl &0.043 & 17.03 &  0.18\\
1992bo &0.018 & 15.42 &  0.21\\
1992bp &0.079 & 18.16 &  0.18\\
1992bs &0.063 & 18.26 &  0.18\\
1993B  &0.071 & 18.40 &  0.18\\
1993O  &0.052 & 17.53 &  0.18\\
1994M  &0.024 & 16.07 &  0.20\\
1994S  &0.016 & 14.83 &  0.22\\
1995ac &0.049 & 17.17 &  0.18\\
1996C  &0.030 & 16.74 &  0.19\\
1996ab &0.125 & 19.47 &  0.19\\
1996bl &0.035 & 16.71 &  0.19\\
\tableline
\end{tabular}
\end{lrbox}
\settowidth{\thiswid}{\usebox{\thisbox}}
\begin{center}
\begin{minipage}{\thiswid}
\caption{SN DATA}
\label{tab:p0ob54}
\usebox{\thisbox}

\end{minipage}
\end{center}
\end{table}

\acknowledgments


\end{document}